# EXPERIMENTS ON A CONDUCTION COOLED SUPERCONDUCTING RADIO FREQUENCY CAVITY WITH FIELD EMISSION CATHODE

Y. Ji, R.C. Dhuley, C. Edwards, J.C.T. Thangaraj, S. Posen
Fermi National Accelerator Labortaory, Batavia, IL, USA
D. Mihalcea, P. Piot O. Mohsen
Department of Physics and Northern Illinois Center for Accelerator & Detector Developement,
Northern Illinois University, DeKalb, IL, USA
I. Salehinia, V. Korampally
Department of Engineering, Northern Illinois University, Dekalb, IL, USA

## Abstract

To achieve Ampere-class electron beam accelerators the pulse delivery rate need to be much higher than the typical photo injector repetition rate of the order of a few kilohertz. We propose here an injector which can, in principle, generate electron bunches at the same rate as the operating RF frequency. A conduction-cooled superconducting radio frequency (SRF) cavity operating in the CW mode and housing a field emission element at its region of high axial electric field can be a viable method of generating high-repetition-rate electron bunches. In this paper, we report the development and experiments on a conduction-cooled $Nb_3Sn$ cavity with a niobium rod intended as a field emitter support. The initial experiments demonstrate $\sim 0.4$ MV/m average accelerating gradient, which is equivalent of peak gradient of 3.2 MV/m. The measured RF cavity quality factor is $1.4 \times 10^8$ slightly above our goal. The achieved field gradient is limited by the relatively low input RF power and by the poor coupling between the external power supply and the RF cavity. With ideal coupling the field gradient can be as high as 0.6 MV/m still below our goal of about 1 MV/m.

## INTRODUCTION

High current (>100 mA) electron beam accelerators have applications ranging from building new radiation sources to medical applications and water treatment in large metropolitan areas. [1] The high charge beam requires continuous wave (CW) operation of the accelerator and electron bunch emission rate close to the operating RF frequency.

Unlike normal conducting copper cavities, superconducting RF (SRF) cavities made of niobium or $Nb_3Sn$ can be operated in the CW mode [2], which makes them an ideal building block of a high-repetition rate field emission electron source. Normally, SRF cavities are operated in liquid helium that provides the cryogenic environment. However, they can also be conduction-cooled with compact, closed-cycle 4 K cryocoolers as demonstrated by Dhuley et al. [3,4]. In this work, Dhuley et al. produced the first-ever practical CW gradients using an SRF cavity without liquid helium [5, 6], making the SRF cavity highly accessible for other applications such as high-repetition-rate field emitters.

Mohsen et al. has developed a multiphysics simulation model for a SRF cavity based field [7]. Using Dhuley et

al.'s conduction-cooled 650 MHz $Nb_3Sn$ cavity and pulse tube cryocooler parameters, Mohsen calculated that a niobium rod inside the cavity with its tip near the center of the cavity can produce peak field of 8 MV/m at its tip. Using Dhuley's conduction-cooled SRF cavity test setup, Mohsen et al.'s initial experiments using a niobium rod placed inside a bulk niobium cavity produced a tip CW gradient of $\sim 0.9$ MV/m. The performance in the experiment was limited by contamination on the stem and the flange. To overcome this limitation, the next set of experiments were conducted using a high-$Q_0$ $Nb_3Sn$ cavity and a surface-cleaned niobium rod. This paper presents the preliminary results from the new configuration and future experiment plans.

## EXPERIMENTAL SETUP

The cavity used in this experiment is a single-cell, 650 MHz, SRF grade niobium (RRR>300) niobium cavity coated with $Nb_3Sn$ on the RF surface. The RF power system provides up to 10 W of CW power at f=650 MHz and $\Delta$f=5 MHz. The cavity is cooled by high purity 5N aluminum thermal link, connected to a cryocooler with 2 W of cooling capacity at 4.2 K.

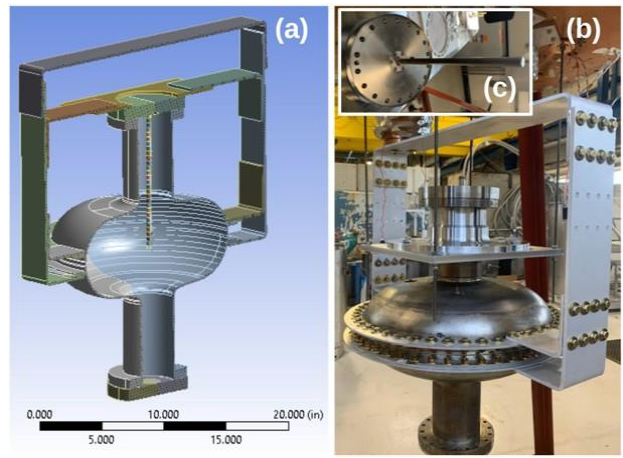

Figure 1: (a) CAD model of the support rod inside the RF cavity. (b) Picture of the RF cavity and (c) the support rod with flange.

Figure 1 displays the rod used to extend the field-emitter into the center of the SRF cavity, where the electric field reaches its maximum. The rod is made of niobium (RRR>300). The rod is attached to the upper flange and the field-emitter array should normally be glued to the open tip of the rod. The desired beam current can be achieved when the peak field at the location of the field-emission array is about 8 MV/m (or about 1 MV/m average field). To ensure that the dissipated power through the SRF cavity walls is within the cryocooler cooling power the Q-factor of the cavity should be in excess of $10^8$. More details are available in [8] and [9]. Though, this experiment does not attach the field-emission array on the tip of the Niobium rod.

## CAVITY BASELINE PERFORMANCE

The $E_{acc}$ and $Q_0$ is measured first for the Nb$_3$Sn coated niobium cavity without the niobium emitter-support rod. This measurement, shown in figure 2, gives the baseline of the cavity performance. Two measurements are shown (a) the first with the natural cooldown of the cryocooler and (b) the second with controlled cooldown of the cryocooler that leads to slow, spatially uniform cooling of the cavity. The latter reduces thermo-current induced trapped magnetic flux in the Nb$_3$Sn layer and consequently produces higher $Q_0$ compared to natural cryocooler cooldown. The highest $E_{acc}$ achieved is ~9.7 MV/m at $Q_0$ of $2 \times 10^{10}$ with controlled cooldown. The temperature of the cavity is shown in the color scale in figure 2. The highest temperature recorded on the cavity is 4.7 K, which is much below superconducting transition temperature of both niobium and Nb$_3$Sn.

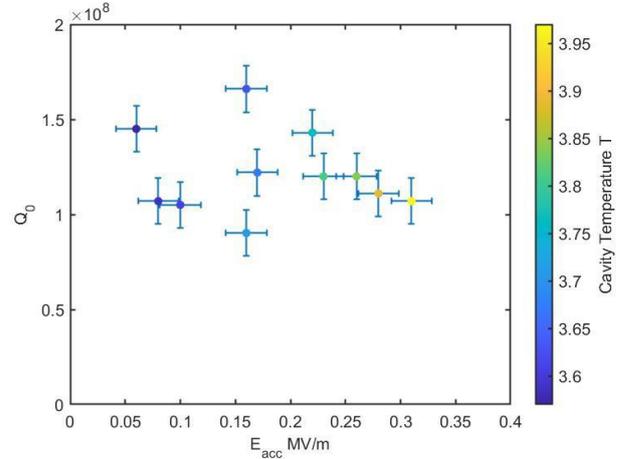

Figure 3: The $Q_0$ vs $E_{acc}$ for the Nb$_3$Sn coated cavity with niobium rod inserted and then natural cooled.

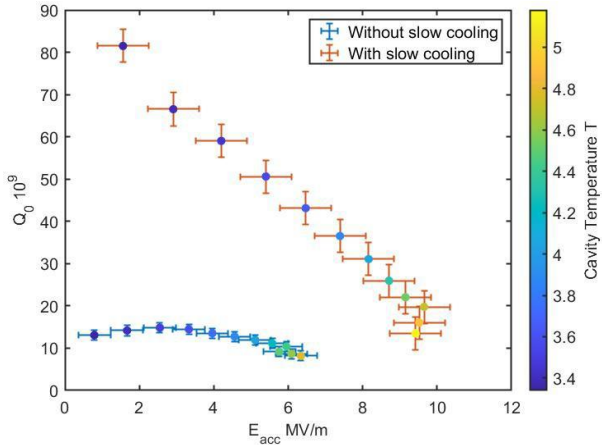

Figure 2: The $Q_0$ vs $E_{acc}$ for the Nb$_3$Sn coated cavity without and with uniform cavity cooling

## CAVITY PERFORMANCE WITH EMISSION ROD

After establishing the baseline for the cavity, the niobium rod was installed (as shown in figure 1) and all tests were performed again. The experiment with the niobium rod was only done with natural cooling, slow cooling experiment was not done.

Figure 3 is the $E_{acc}$ and $Q_0$ for the cavity with the superconductive rod. The achieved $E_{acc}$ is ~0.4 MV/m and $Q_0$ is $1.4 \times 10^8$. The $Q_0$ is three orders of magnitude lower than the empty cavity. Figure 4 shows the data measured with the niobium rod. Figure 4 (a) displays dissipated, reflected and transmitted powers. Figure 4 (b) shows temperature at the cryocooler. Figure 4 (c) and (d) shows the $Q_0$ and $E_{acc}$ as measured and as calculated based on the BCS theory surface resistance added with 10 n$\Omega$ residual resistance (the shaded areas).

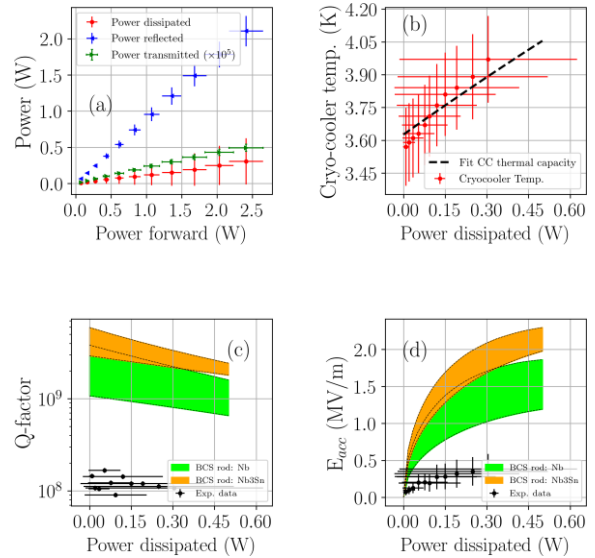

Figure 4: Measured dissipated, reflected and transmitted-https://www.overleaf.com/project/622639e08a4c7f810a745440 powers (a) Measured temperatures at cryocooler (b) Measured RF cavity Q-factor as a function of dissipated power (c) Average accelerating field (d).

Because the introduction of the rod greatly decreased the coupling strength, the $E_{acc}$ is substantially lower than the accelerating gradient achieved with the empty cavity. Figure 4 (a) and (b) demonstrated the power measured and the temperature with respect to dissipated power. Figure 4 (c) and (d) demonstrated the $Q_0$ factor and $E_{acc}$ based on power dissipation. In figure 4, the experiment is also compared with theoretical prediction based on the BCS theory (green plot) and theoretical prediction based on potential equipment improvements.

The experimental data is compared with BCS predictions. In figure 4 (c) and (d), BCS prediction is calculated (green region). The range of the BCS prediction is based on the rod tip temperature $T_{ti\ p}$. Since $T_{ti\ p}$ is not directly measured the upper and lower limits of the BCS calculations are based on two observations: $T_{ti\ p} > T_{f\ lange}$ to evaluate the upper limit of $Q_0$ and $T_{ti\ p} < T_c = 9.2$ K to evaluate the lower limit of $Q_0$. The upper bound of the BCS prediction is calculated based on the upper flange temperature. A potential improvement to current set-up is to coat the rod and the flange with $Nb_3Sn$. In figure 4, theoretical prediction of the coating is demonstrated in the orange region. Although the temperature of the system used in the calculation is based on pure niobium rather than $Nb_3Sn$ coated niobium, the calculation demonstrated that the $Nb_3Sn$ coating will give a potential 40% increase in the $Q_0$ factor.

Because of the low coupling, both power dissipation and stored energy in the cavity are low. The energy stored in the cavity is calculated based on

$$U = \frac{4\beta P_f\ Q_{cav}}{(\beta + 1)^2\ \omega} \quad with\ \beta = \frac{Q_{cou\ pler}}{Q_{cav}} \tag{1}$$

where $P_f$ is the forward power at the cavity input. The measurement gives $\beta_{in} \approx 0.03$. With such low coupling, both power dissipation and stored energy in the cavity are low, which leads to lower accelerating gradient. Better coupling is expected to improve the accelerating gradient produced by the cavity.

The average accelerating gradient achieved is 0.4 MV/m, with corresponding peak field at 3.2 MV/m at the tip of the rod. Based on eq 1, with ideal coupling, the accelerating gradient can increase by a factor of 2.85. This increase will be partially offset by the increase of power dissipation (about 2 W). The increase of the power dissipation will increase the cavity temperature by 1.2 K. Based on BCS theory calculation, the $Q_0$ will decrease by a factor of 3.4 because of the temperature increase. Eventually, the ideal coupling could lead to 0.6 MV/m accelerating gradient, which equates to 4.8 MV/m peak field at the rod tip.

## CONCLUSION AND FUTURE PLAN

In this paper, we have reported the first measurements of $E_{acc}$ on an SRF cavity a niobium rod intended to be the support for a field-emission cathode. Without the rod, the system achieved $E_{acc}$ of 9.7 MV/m at $Q_0$ of $2 \times 10^{10}$. With the rod, the cavity achieved $Q_0 = 1.4 \times 10^8$ and $E_{acc} = 0.4$ MV/m. This is equivalent to peak field of 3.2 MV/m at the tip of the niobium rod Although the configuration produced measurable gradient, the current performance is limited by the input coupling at the cavity. Based on calculation of the ideal coupling, the system could potentially achieve $E_{acc} = 0.6$ MV/m, which equates to 4.8 MV/m peak field at the rod tip. The system performance could further be improved by coating the rod and the upper flange with $Nb_3Sn$. Based on our calculations, by coating the niobium rod and flange with $Nb_3Sn$, the cavity quality factor $Q_0$ increases by a factor of 3 and the electric field by about 40%.

## ACKNOWLEDGMENT

This manuscript has been authored by Fermi Research Alliance, LLC under Contract No. DE-AC02-07CH11359 with the U.S. Department of Energy, Office of Science, Office of High Energy Physics. This work was supported by the US Department of Energy (DOE) under contract DE-SC0018367 with Northern Illinois University.